\documentclass{ws-ijmpc}

\usepackage{algorithm,algorithmic}
\usepackage[normalem]{ulem}

\begin{document}

\markboth{Authors' Names}
{Instructions for Typing Manuscripts (Paper's Title)}

%
\catchline{}{}{}{}{}
%

\title{Community detection in bipartite networks using weighted symmetric binary matrix factorization}

\author{Zhong-Yuan Zhang}

\address{School of Statistics and Mathematics, Central University of Finance and Economics,\\ Beijing, China\\
zhyuanzh@gmail.com }

\author{Yong-Yeol Ahn}

\address{School of Informatics and Computing, Indiana University Bloomington, IN, USA\\
yyahn@indiana.edu}

\maketitle

\begin{history}
\received{Day Month Year}
\revised{Day Month Year}
\end{history}

\begin{abstract}
In this paper we propose weighted symmetric binary matrix factorization (wSBMF)
framework to detect overlapping communities in bipartite networks, which describe relationships between two types of nodes.
Our method improves performance by recognizing the distinction between two types of missing edges---ones among the nodes in each node type and the others between two node types.
Our method can also explicitly assign community membership and distinguish outliers from overlapping
nodes, as well as incorporating existing knowledge on the network. We propose a generalized partition density for
bipartite networks as a quality function, which identifies the most appropriate number of communities. The
experimental results on both  synthetic and  real-world networks
demonstrate the effectiveness of our method.
\end{abstract}


\section{Introduction}\label{intro}

Community structure is a common characteristic of various complex networks
found in biological, social, and information systems,
etc.~\cite{Girvan02,newman2004finding,protein,newman2006modularity,linkcommunity,review3,review2,review1}. A community is commonly defined as a
densely interconnected set of nodes that is loosely connected with the
rest of the network~\cite{Girvan02}. Studies have shown that community
structures are highly relevant to the organization and  functions of the
network. For instance, communities in social networks correspond to social
circles~\cite{Girvan02}; communities in protein-protein interaction networks
capture functional modules~\cite{linkcommunity,protein}; and communities affect the
spread of behaviors and ideas~\cite{protein,spread,virality}.

Although numerous community detection methods have been proposed, relatively
few methods are designed for bipartite networks~\cite{bipartitemodularity,overlapping,evolutionary,biclique,liu2009,lind2005cycles,lind2007new}.  A bipartite network
$G(\Delta,\Gamma,E)$ contains two disjoint types of nodes, $\Delta$ and
$\Gamma$, and the edge set $E$ connecting the two parts. There is no
edge among vertices in $\Delta$ and among those in $\Gamma$. Many systems can
be naturally modelled as bipartite networks~\cite{biclique,flavor}.  For
instance, a metabolic network can be considered as a bipartite network of
reactions and metabolites~\cite{jeong}. Many unipartite networks are derived
from bipartite ones. For instance, a scientific collaboration network is
derived from an author-paper bipartite network~\cite{collaboration}. A community
in a bipartite network $G(\Delta,\Gamma,E)$ can be defined as a set of
nodes --- from both $\Delta$ and $\Gamma$ --- that are densely interconnected.
Bipartite community detection is not necessarily equivalent to unipartite
community detection on the projected networks, because the projection often
destroys important information~\cite{zhou2007bipartite,overlapping,biclique}. Here we would like to point out the difference between the missing edge among $\Delta$ and among $\Gamma$, and that between $\Delta$ and $\Gamma$.
Imagine a network of people
and their affiliations. With  complete information
about people's affiliation, the absence of edge $(i,\,j)$
$(i \in \Delta,\, j \in \Gamma)$ means
that the person $i$ does not belong to the organization $j$. However, the absence of edge $(i,k)
\, (i,\,k \in \Delta)$ simply indicates that we do not know
the direct social relationships between $i$ and $k$.

In our previous work we  proposed the Symmetric Binary Matrix Factorization (SBMF) to detect overlapping communities in unipartite networks and
demonstrated its effectiveness~\cite{SBMF}. In this paper, we propose weighted Symmetric Binary Matrix Factorization
model to detect overlapping communities in bipartite networks. The model can differentiate between the two kinds of missing edges in the bipartite network to improve detecting performance. The model allows
us explicitly to  assign community membership to nodes and distinguish outliers
from overlapping nodes while providing a way to analyze the strength of
membership and incorporate
existing information. To quantify the goodness of the communities that we
found, we generalize partition density and use it to select the most
appropriate number of communities.

\section{Methods}
\label{sec:1}

\subsection{Weighted Symmetric Binary Matrix Factorization}
The adjacency matrix of an undirected and unweighted simple graph $G$ with $n$ nodes can be defined as:
$$
A_{ij} = \left\{\begin{array}{rcl}
         1, & & \mbox{if}\ i\sim j\\
         0, & & \mbox{if}\ i = j\ \mbox{or}\ i\nsim j,\\
      \end{array}\right.
$$
where $i\sim j$ means there is an edge and $i\nsim j$ means there is no edge.

 Imagine an unweighted and undirected bipartite network
$G(\Delta,\Gamma,E)$, which has $n_\Delta$ and $n_\Gamma$ nodes in $\Delta$ and
$\Gamma$, respectively, and an edge set $E$ connecting the two parts. The
corresponding adjacency matrix $A$ can be split into four blocks after the
$n_\Delta$th row and the $n_\Delta$th column:
$$
A = \left[ \begin{array}{c c}\vspace{2mm}
 \mathbf{0_\Delta} & B \\
              B^T  & \mathbf{0_\Gamma}
              \end{array}
\right],
$$
where $\mathbf{0_\Delta}$ and $\mathbf{0_\Gamma}$ are null matrices
of size $n_\Delta\times n_\Delta$ and $n_\Gamma\times n_\Gamma$, respectively, \uline{and}
$$B_{ij} = \left\{\begin{array}{rcl}
         1, & & \mbox{if}\ i\sim j,\ i\in\Delta,\, j\in\Gamma\\
         0, & & \mbox{if}\ i\nsim j,\ i\in\Delta,\, j\in\Gamma\\
      \end{array}\right.
$$
The meaning of the zeros in $\mathbf{0_\Delta}$, $\mathbf{0_\Gamma}$ is different from that in $B$.
If $B$ captures all existing connections perfectly, then all zeros in $B$ indicate the absence of the corresponding edges. By contrast, the zeros in $\mathbf{0_\Delta}$ and $\mathbf{0_\Gamma}$ represent missing information, rather than the absence of edges.
To use this information,
we introduce a weight matrix $L$ of size $n\times n$ to handle these unobserved
or missing values~\cite{lee2010semi}, which can be defined as:
$$
 \begin{array}{rcl}
 L_{ij} & = & \left\{
               \begin{array}{rl}
               \gamma & \mbox{if $A_{ij}$ is observed} \\
               0      & \mbox{if $A_{ij}$ is unobserved},
               \end{array}
               \right.
 \end{array}
$$
where $\gamma$ is a nonnegative weight parameter that captures the reliability of $A_{ij}.$
 For standard bipartite networks, $L$ can be formulated as: $$L=\left[ \begin{array}{c c}\vspace{2mm}
\mathbf{0_\Delta} & \mathbf{I_{\Delta,\, \Gamma}} \\
\mathbf{I_{\Gamma,\, \Delta}}  & \mathbf{0_\Gamma}
\end{array} \right],$$ where $\mathbf{I_{\Delta,\, \Gamma}}$ and $\mathbf{I_{\Gamma,\, \Delta}}$ are matrices where all entries are one, meaning that only the zeros in $B$ are considered. The sizes of $\mathbf{I_{\Delta,\, \Gamma}}$ and $\mathbf{I_{\Gamma,\, \Delta}}$ are $n_\Delta\times n_\Gamma$ and $n_\Gamma\times n_\Delta$, respectively.

Our weighted Symmetric Binary Matrix Factorization (wSBMF) model can be defined as the following constrained nonlinear programming:
\begin{equation}
\begin{array}{cc}\label{Eq:04}\vspace{1mm}
\min\limits_{U} & \hspace{-2mm} \|L\circ(A-UU^T)\|_1+\sum\limits_i(1-\Theta(\sum\limits_jU_{ij}))\\
\hspace{-1mm}\mbox{subject to} & U_{ij}^2-U_{ij}=0,\, i = 1,2,\ldots,n,\, j=1,2,\ldots,c,
\end{array}
\end{equation}
where $\circ$ represents element-wise multiplication (Hadamard product); $A$ is the adjacency matrix of size $n\times n$ ($n=n_\Delta+n_\Gamma$); $U$ is the community membership matrix such that $U_{it}=1$ if node $i$ is in the community $t$, and $0$ if otherwise; 
Note that numerical experiments show
that the Frobenius norm on the sparse adjacency matrix $A$ often results in the
ultra-sparsity of $U$, even null matrix $U$, which is not informative enough for real analysis.
We use 1-norm instead to obtain more reasonable and explainable matrix $U$.
1-norm of a matrix $X$ is the largest column sum of $\mbox{abs}(X)$, where $\mbox{abs}(X)_{ij}=\mbox{abs}(X_{ij})$, and $\mbox{abs}(\cdot)$ is the absolute value; $\Theta$ is the Heaviside step function such that for some matrix $X$,
$$
\Theta(X)_{ij}:=\left\{\begin{array}{rl}\vspace{2mm}
                  1 & \mbox{if}\, X_{ij}>0;\\
                  0 & \mbox{if}\, X_{ij}\leqslant0.
                 \end{array}
                 \right.
$$
 $L$ chooses which entries of the adjacency matrix should be considered in the optimization and thus allows us to incorporate existing knowledge. For instance, if we already know that some edges are present between nodes in $\Delta$, then we can update the corresponding elements of $L$ from zero to $\gamma$. If we want to ignore edges in $B$, we can simply update the corresponding element of $L$ from one to zero. We can even vary $\gamma$ across elements if we can assess the reliability of the incorporated knowledge. 

We initialize $U$ by solving the following weighted Symmetric Nonnegative Matrix Factorization  model:
\begin{equation}
\begin{array}{rl}\label{Eq:02}\vspace{1mm}
\min\limits_{U} & \|L\circ(A-UU^T)\|_F^2\\\vspace{3mm}
\mbox{subject to} & U_{ij}\geqslant0,\, i = 1,2,\ldots,n,\, j=1,2,\ldots,c,\\
     & \sum_{j=1}^cU_{ij}=1, \, i=1,2,\ldots,n.
\end{array}
\end{equation}
Then we fix $U$, and discretize the domain $\{u: 0\leqslant u\leqslant \max(U)\}$ to find $\hat{u}$ that minimizes the following, simpler optimization problem:
 \begin{equation}
\begin{array}{rl}\label{Eq:03}\vspace{1mm}
\min\limits_{U} & \|L\circ(A-\Theta(U-u)\Theta(U-u)^T)\|_1+\\
                & \hspace{25mm}+\sum\limits_i(1-\sum\limits_j\Theta(U-u)_{ij})
\end{array}
\end{equation}
where $u$ is a scalar. Finally, we obtain the binary matrix $U$ as follows: $$U: =\Theta(U-\hat{u}).$$

To optimize $U$ for model (\ref{Eq:02}), we initialize $U$ using the algorithm of alternative least squares error developed for NMF \cite{1994positive,berry2007algorithms}:
\begin{equation}
\begin{array}{rl}\label{Eq:01}\vspace{1mm}
\min\limits_{U_1, U_2} & \displaystyle\|B-U_1U_2^T\|^2_F\\
\mbox{subject to} & U_{1}\geqslant0,\, U_{2}\geqslant0.
\end{array}
\end{equation}
See Appendix: Algorithm 1.

Then, based on the boundedness theorem \cite{BMF,BMF2,chapter}, we normalize $U_1$ and $U_2$ to balance their scales:
\begin{equation}\label{Eq:05}
U_1 = U_1D_1^{-1/2}D_2^{1/2},\ \ \ U_2 = U_2D_2^{-1/2}D_1^{1/2}
\end{equation}
where
$$
\begin{array}{rcl}
\hspace{-2mm}D_1 & = & \mbox{diag}\left(\max U_1(:,1),\max U_1(:,2),\cdots,\max U_1(:,c)\right);\\
\hspace{-2mm}D_2 & = & \mbox{diag}\left(\max U_2(:,1),\max U_2(:,2),\cdots,\max U_2(:,c)\right);
\end{array}
$$
and $\mbox{diag}(a_1,a_2,\ldots,a_n)$ is the diagonal matrix whose diagonal entries starting from the upper left corner are $a_1, a_2,\ldots, a_n.$ $U_1(:,i)$ is the $i$th column of $U_1$. Finally, we merge $U_1$ and $U_2$ into $U$ such that $U = \left[ \begin{array}{c}\vspace{2mm}
U_1\\
U_2
\end{array} \right]$, and employ the algorithm of multiplicative update rules for model (\ref{Eq:02}). See Appendix: Algorithm 2.

\subsection{Model Selection}
We have proposed a modified partition density
to select the appropriate number of communities~\cite{linkcommunity,SBMF}. The modified partition density is defined as:
$$
  D = \sum_{\alpha=1}^c\frac{1}{q^{(\alpha)}}\frac{n^{(\alpha)}}{N}D^{(\alpha)},
$$
where $D^{(\alpha)}$ is the partition density of community $\alpha:$
$$
   D^{(\alpha)} = \frac{m^{(\alpha)}-\underline{m}^{(\alpha)}}{\overline{m}^{(\alpha)}
   -\underline{m}^{(\alpha)}},
$$
and $\underline{m}^{(\alpha)}=(n^{(\alpha)}-1),$ $\overline{m}^{(\alpha)}=n^{(\alpha)}(n^{(\alpha)}-1)/2$ are the minimum and maximum possible numbers of links between the nodes in the community $\alpha$, respectively;  $n^{(\alpha)}$ and $m^{(\alpha)}$ are the number of nodes and the number of edges in the community $\alpha$, respectively; $q^{(\alpha)}=\max_{j\in \alpha}l_j$ is the maximum number of community memberships ($l_j$) among the nodes ($j$) that belong to the community $\alpha$; $N$ is the sum of the sizes of different communities and the number of outliers.

Here we generalize it for bipartite networks by transforming each bipartite
community to a unipartite one and getting the corresponding partition density. For a community $\alpha,$ we define the subnetwork $G^{(\alpha)}$
as the set of nodes in $\alpha$ and the edges among them.  The subnetwork has
$n_\Delta^{(\alpha)}$ nodes in $\Delta$ and $n_\Gamma^{(\alpha)}$ nodes in
$\Gamma$, and the corresponding adjacency matrix is
$$ A^{(\alpha)}= \left[
\begin{array}{c c} \mathbf{0} & B^{(\alpha)} \\ B^{(\alpha)T}  & \mathbf{0}
\end{array} \right].
$$
%
%
Then we transform the bipartite subnetwork $G^{(\alpha)}$ to a unipartite
subnetwork $G^{(\alpha)'}$ by overlaying the two projections onto $\Delta$ and
$\Gamma$. The adjacency matrix $A^{(\alpha)}$ becomes: $$ A^{(\alpha)'}= \left[
\begin{array}{c c} B^{(\alpha)} B^{(\alpha)T} & B^{(\alpha)} \\ B^{(\alpha)T} &
B^{(\alpha)T} B^{(\alpha)} \end{array} \right], $$ and the diagonal elements
indicate the number of neighbors in the other part that the corresponding node
has. The values of $m^{(\alpha)}, \overline{m}^{(\alpha)}$, and
$\underline{m}^{(\alpha)}$ are changed to: $$ \begin{array}{l}
\hspace{-6mm}m^{(\alpha)'}  =
\sum_{i,j}(A^{(\alpha)'}-\mbox{diag}(A^{(\alpha)'}))_{ij}/2, \end{array} $$
where $\mbox{diag}(A^{(\alpha)'})$ is the diagonal matrix whose diagonal
entries are those of $A^{(\alpha)'}$; $$ \begin{array}{l}\vspace{3mm}
\overline{m}^{(\alpha)'}=\\\vspace{3mm}
\left[\displaystyle\frac{n_\Delta^{(\alpha)}
(n_\Delta^{(\alpha)}-1)}{2}n_\Gamma^{(\alpha)}
+\displaystyle\frac{n_\Gamma^{(\alpha)}(n_\Gamma^{(\alpha)}-1)}{2}
n_\Delta^{(\alpha)}
 +n_\Delta^{(\alpha)} n_\Gamma^{(\alpha)} \right];
\end{array} $$ and $$ \begin{array}{l}\vspace{3mm} \underline{m}^{(\alpha)'}=
\left[(n_\Delta^{(\alpha)}-1)
+(n_\Gamma^{(\alpha)}-1)
+(n_\Delta^{(\alpha)}+n_\Gamma^{(\alpha)}-1)\right].
\end{array} $$ Then $D^{(\alpha)}$ becomes: $$
D^{(\alpha)'} = \frac{m^{(\alpha)'}-\underline{m}^{(\alpha)'}}
{\overline{m}^{(\alpha)'}-\underline{m}^{(\alpha)'}}, $$ and the generalized
partition density is: $$ D' =
\sum_{\alpha=1}^c\frac{1}{q^{(\alpha)}}\frac{n^{(\alpha)}} {N}D^{(\alpha)'}. $$

\subsection{An illustrative Example}
\begin{figure*}[!ht]
\begin{center}
 \includegraphics[height=50mm,width=155mm]{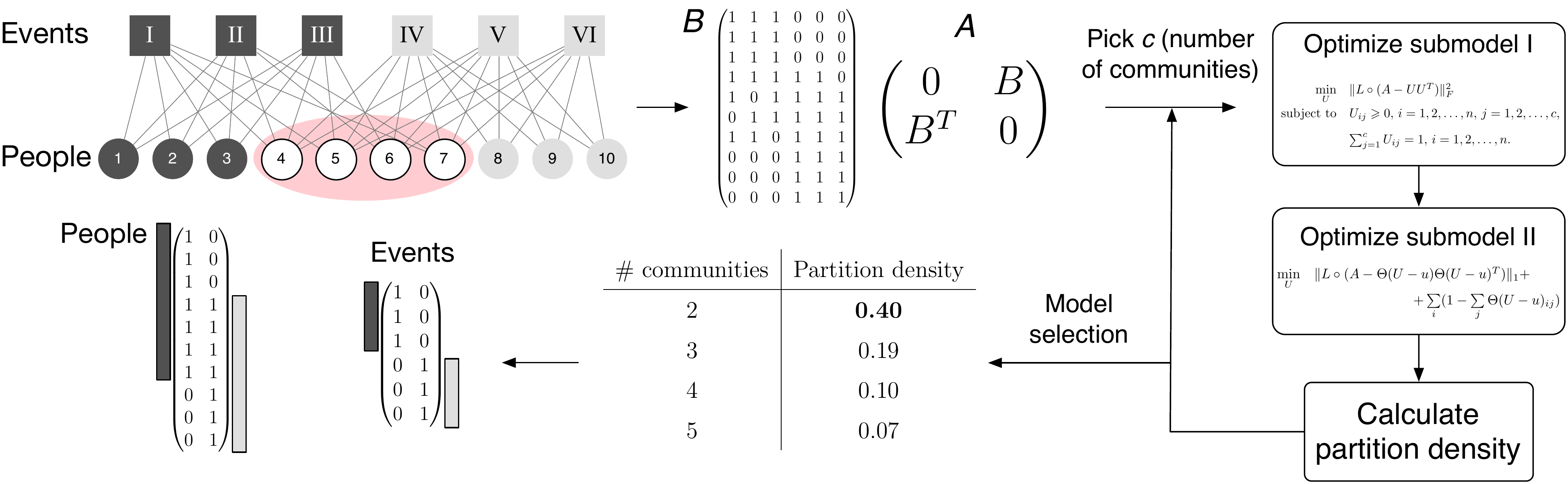}
\end{center}
\caption{{\bf Illustration of wSBMF method.} The network consists of events and people and exhibits two overlapping
groups where some individuals (4-7) belong to both communities.}\label{Fig:01}
\end{figure*}

We show a small example that illustrates how the method works. Figure
\ref{Fig:01} exhibits a bipartite network with two communities, which can be
clearly recovered by our approach.
Specifically, for $c=2$ we have $m^{(1)}=136, m^{(2)}=114$;
$\underline{m}^{(1)}=35, \underline{m}^{(2)}=35;$ $\overline{m}^{(1)}=147,
\overline{m}^{(2)}=147$; $q^{(1)}=2, q^{(2)}=2;$ and $N=20.$ Let us
illustrate how we can incorporate existing knowledge. If
we know that Nodes III and IV are in the same community, then we can revise $A$
and $L$ such that the elements in the positions of $(13,14)$ and $(14,13)$ are
1. The result for events is changed to $$%
\left[
\begin{array}{c c c c c c}
1 & 1 & 1 & 1 & 1 & 0\\
0 & 0 & 0 & 1 & 1 & 1\\ \end{array}
\right]^T, $$%
which group III and IV together.

Note that the bipartite network can be projected onto the Event part or onto the People part. Two events are connected if they have at least one common neighbor in the People part, resulting in a complete network containing six nodes. The loss of information is obvious and the community structures vanish, which means that the problem of community detection in bipartite networks is not reducible to unipartite case.
\subsection{Possible Extensions}
The wSBMF model can be naturally extended to $M$-partite networks, whose adjacency matrix can be split into $M\times M$ blocks:
$$
A = \left[ \begin{array}{c c c c c}\vspace{2mm}
        \mathbf{0}_{\Lambda_1,\Lambda_1} & B_{{\Lambda_1,\Lambda_2}} & \mathbf{0}_{\Lambda_1,\Lambda_3} & \cdots & \mathbf{0}_{\Lambda_1,\Lambda_{M}}\\
        B_{{\Lambda_1,\Lambda_2}}^T  & \mathbf{0}_{\Lambda_2,\Lambda_2} & B_{{\Lambda_2,\Lambda_3}} & \cdots & \mathbf{0}_{\Lambda_2,\Lambda_{M}}\vspace{2mm}\\
        \mathbf{0}_{\Lambda_3,\Lambda_1} & B_{{\Lambda_2,\Lambda_3}}^T & \mathbf{0}_{\Lambda_3,\Lambda_3} & \cdots & \mathbf{0}_{\Lambda_3,\Lambda_M} \vspace{2mm}\\
        \vdots & \cdots & \cdots & \cdots & \vdots\vspace{2mm}\\
        \vdots & \cdots & \cdots & \cdots & \vdots\vspace{2mm}\\
        \mathbf{0}_{\Lambda_{M-1},\Lambda_{1}} & \mathbf{0}_{\Lambda_{M-1},\Lambda_2} & \cdots & \mathbf{0}_{\Lambda_{M-1},\Lambda_{M-1}} & B_{{\Lambda_{M-1},\Lambda_M}} \vspace{2mm}\\
        \mathbf{0}_{\Lambda_M,\Lambda_{1}} & \mathbf{0}_{\Lambda_M,\Lambda_2} & \cdots & B_{{\Lambda_{M-1},\Lambda_M}}^T & \mathbf{0}_{\Lambda_M,\Lambda_M}
              \end{array}
\right],
$$
where $\mathbf{0}_{\Lambda_i,\Lambda_j}$ is null matrix of size $n_{\Lambda_i}\times n_{\Lambda_j},$ and
$$
B_{\Lambda_i,\Lambda_{i+1} ab} = \left\{\begin{array}{rcl}
         1, & & \mbox{if}\ a\sim b,\ a\in\Lambda_i,\, b\in\Lambda_{i+1}\\
         0, & & \mbox{if}\ a\nsim b,\ a\in\Lambda_i,\, b\in\Lambda_{i+1},\ \ i=1,2,\cdots,M-1.\\
      \end{array}\right.
$$
In this case, $L$ should be reformulated as:
$$
L = \left[ \begin{array}{c c c c c}\vspace{2mm}
        \mathbf{0}_{\Lambda_1,\Lambda_1} & \mathbf{I}_{{\Lambda_1,\Lambda_2}} & \mathbf{0}_{\Lambda_1,\Lambda_3} & \cdots & \mathbf{0}_{\Lambda_1,\Lambda_{M}}\\
        \mathbf{I}_{{\Lambda_2,\Lambda_1}}  & \mathbf{0}_{\Lambda_2,\Lambda_2} & \mathbf{I}_{{\Lambda_2,\Lambda_3}} & \cdots & \mathbf{0}_{\Lambda_2,\Lambda_{M}}\vspace{2mm}\\
        \mathbf{0}_{\Lambda_3,\Lambda_1} & \mathbf{I}_{{\Lambda_3,\Lambda_2}} & \mathbf{0}_{\Lambda_3,\Lambda_3} & \cdots & \mathbf{0}_{\Lambda_3,\Lambda_M} \vspace{2mm}\\
        \vdots & \cdots & \cdots & \cdots & \vdots\vspace{2mm}\\
        \vdots & \cdots & \cdots & \cdots & \vdots\vspace{2mm}\\
        \mathbf{0}_{\Lambda_{M-1},\Lambda_{1}} & \mathbf{0}_{\Lambda_{M-1},\Lambda_2} & \cdots & \mathbf{0}_{\Lambda_{M-1},\Lambda_{M-1}} & \mathbf{I}_{{\Lambda_{M-1},\Lambda_M}} \vspace{2mm}\\
        \mathbf{0}_{\Lambda_M,\Lambda_{1}} & \mathbf{0}_{\Lambda_M,\Lambda_2} & \cdots & \mathbf{I}_{{\Lambda_M,\Lambda_{M-1}}} & \mathbf{0}_{\Lambda_M,\Lambda_M}
              \end{array}
\right],
$$
where $\mathbf{I}_{{\Lambda_i,\Lambda_j}}$ is matrix where all entries are one with size $n_{\Lambda_i}\times n_{\Lambda_j}$.

\section{Results}
In this section we evaluate the performance of our method using both synthetic
and real-world networks.

\subsection{Datasets Description}
\label{synthetic}
We first
discuss the existing bipartite benchmark networks~\cite{bipartitemodularity}. The benchmark
has five
communities, each having the same number of nodes.
Edges only exist between $\Delta$ and $\Gamma$ with possibility $p_{in}$ if
they are in the same community and $p_{out}$ if otherwise. Often, $p_{in}$ is
set equal to either $0.5$ or $0.9$ and $p_{out}$ is set as $\alpha p_{in}$,
where $\alpha$ varies from $0$ to 1. With increasing $\alpha$, the community
structure becomes less clear. Here we propose two new, more realistic benchmark
graphs that exhibit overlaps, variable community sizes, and fixed density with
different mixing parameters.

\begin{itemize}

\item Non-overlapping communities: This class of networks has four communities
with the same number of nodes (each with 32 from $\Delta$ and 32 from
$\Gamma$). Edges exist only between $\Delta$ and $\Gamma$. On average, each
node has $Z_{in}+Z_{out}=16$ edges. In other words, each node in $\Delta$ has
$Z_{in}$ neighbors within its own community and $Z_{out}$ ones outside. With
decreasing $Z_{out}$, the community structures become clearer.

\item Overlapping communities: This class of networks has $c$ communities and
the number of nodes in each community can differ from each other. A community
$\alpha$ contains $n_\Delta^{(\alpha)}$ nodes and $n_\Gamma^{(\alpha)}$ ones in
$\Delta$ and $\Gamma$ respectively. On average each $\Delta$ node in the
community $\alpha$ has $Z_{in}^{(\alpha)}$ $\Gamma$ neighbors in its own
community and $Z_{out}^{(\alpha)}$ $\Gamma$ neighbors in other communities.
Actually, since we should have
$Z_{in}^{(\alpha)}/n_\Gamma^{(\alpha)}=Z_{in}^{(\alpha')}/n_\Gamma^{(\alpha')},$
and $Z_{out}^{(\alpha)}/(\sum_tn_\Gamma^{(t)}-n_\Gamma^{(\alpha)})=
Z_{in}^{(\alpha')}/(\sum_tn_\Gamma^{(t)}-n_\Gamma^{(\alpha')}), \ \alpha,\,
\alpha'=1,2,\ldots c$, it is enough only to give $Z_{in}^{(1)}$ and
$Z_{out}^{(1)}$ to generate the network. In our setting there are four
communities containing $32$ $\Delta$ nodes and $32$ $\Gamma$ ones in each
community. In addition, there are $t$ overlapping $\Delta$ nodes between
communities $\alpha$ and $\alpha+1$, $\alpha=1,\,2,\,3$.  $Z_{in}^{(1)}$ and
$Z_{out}^{(1)}$ are set to $10$ and $6$, respectively.
\end{itemize}

We also use real-world networks for evaluation.
\begin{itemize}

\item Southern women network~\cite{davis}: This dataset is the network
describing the relations between 18 women and 14 social events.  Edges only
exist between the women and the events, which makes the graph bipartite. There
are 89 edges. The network is commonly used as a benchmark for bipartite community
detection.

\item Senator network\footnote{http://www.senate.gov/}: This is the network of
110 US senators connected by voting records for 696 bills.  There is an edge
    between the senator and the bill if the senator voted for the bill.  We
    remove inactive senators who abstained from more than thirty percent of the
    bills and also the inactive bills which are waived by more than thirty
    percent of senators. The final dataset contains 96 senators and 690
    bills.  There are still abstention cases in
    the network, which are considered as missing values and can be handled by
    $L$.

\end{itemize}

\subsection{Assessment Standards} \label{measure}

Normalized mutual information is used as the standard to evaluate
community structure detection performance. The value can be formulated as
follows~\cite{strehl2003cluster}: $$I_{norm}(M_1, M_2) =
\frac{\sum\limits_{i=1}^{c}
\sum\limits_{j=1}^{c}n_{ij}\ln\frac{n_{ij}n}{n_i^{(1)}n_j^{(2)}}}
{\sqrt{\left(\sum\limits_{i=1}^{c}n_i^{(1)}\ln\frac{n_i^{(1)}}{n}\right)
\left(\sum\limits_{j=1}^{c}n_j^{(2)}\ln\frac{n_j^{(2)}}{n}\right)}}, $$ where
$M_1$ and $M_2$ are the true cluster label and the computed cluster label,
respectively; $c$ is the community number; $n$ is the number of nodes; $n_{ij}$
is the number of nodes in the true cluster $i$ that are assigned to the
computed cluster $j$; $n_i^{(1)}$ is the number of nodes in the true cluster
$i$; and $n_j^{(2)}$ is the number of nodes in the computed cluster $j$.  The
larger the values of NMI, the better the graph partitioning results.  For
overlapping benchmarks we use the generalized normalized mutual information
{\cite{gnmi}}.

\subsection{Results}
\label{simulation}
We compare our method with the BRIM model \cite{bipartitemodularity}, which is the only method that we can get the codes, on the
synthetic benchmarks. Note the the BRIM method cannot handle overlapping communities and missing values in the network. To show that the problem of detecting overlapping communities in bipartite networks is not trivial and cannot be reduced
to the unipartite case, we also compare our method with SBMF model \cite{SBMF} on the two unipartite networks $\Delta$ and $\Gamma$, where the two nodes are connected if they have at least one common neighbor.

In many real scenarios there is background information available. We can
incorporate it into the detection process by revising the objective matrix $A$
and the weight matrix $L$ to improve the performance of detection and the
interpretability of the results. Specifically,  we consider two types of
background information for node pairs of the same type (i.e., $\Delta$ or $\Gamma$): (i)
\texttt{existence constraint} $C_e$: $(i,j)\in C_e$ means that nodes $i$ and
$j$ are connected; (ii) \texttt{absence
constraint} $C_{a}$: $(i,j)\in C_{a}$ means that nodes $i$ and $j$ are not connected.

We only consider incorporating background information on the nodes in $\Delta$
in this paper for simplicity.  Given a bipartite network with $n_\Delta$ nodes
in $\Delta$, there are $n_\Delta(n_\Delta-1)/2$ pairs of nodes available. We
randomly select five percent of pairs for prior information: if the two nodes
in one pair have the same community label, we assume that they belong to
$C_{e}$, otherwise they belong to $C_{a}$ \cite{semisupervised,massive}.
The zero matrices $\mathbf{0}_\Gamma$ in $A$ and $L$ are revised accordingly:
\begin{equation}
\mathbf{0}_{\Delta\, ij} = \left\{\begin{array}{rcl} 1, & & \mbox{if}\ (i,\,
j)\in C_{e}\\ 0, & & \mbox{otherwise}, \end{array}\right.
\end{equation}
where $\mathbf{0}_\Delta$ is the submatrix in $A$.  \begin{equation}
\mathbf{0}_{\Delta\,ij} = \left\{\begin{array}{rcl} \gamma, & & \mbox{if}\
(i,\, j)\in C_{e}\ \mbox{or}\ (i,\, j)\in C_{a}\\ 0, & & \mbox{otherwise},
\end{array}\right.  \end{equation} where $\mathbf{0}_\Delta$ is the submatrix
in $L$. We set $\gamma$ equal to $1$. 

\begin{figure}[!ht]
\begin{center}
\includegraphics[height=53mm,width=73mm]{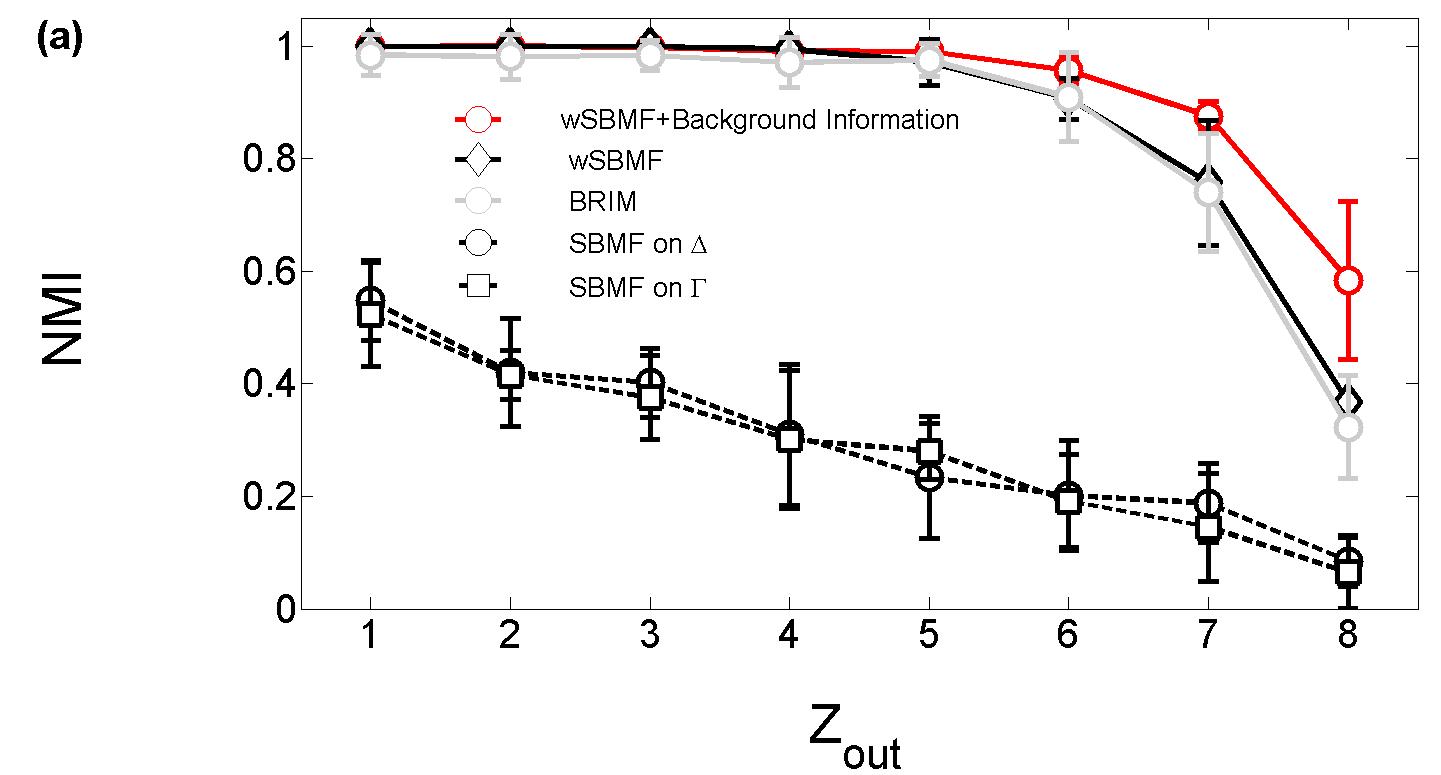}\vspace{5mm}
\includegraphics[height=50mm,width=73mm]{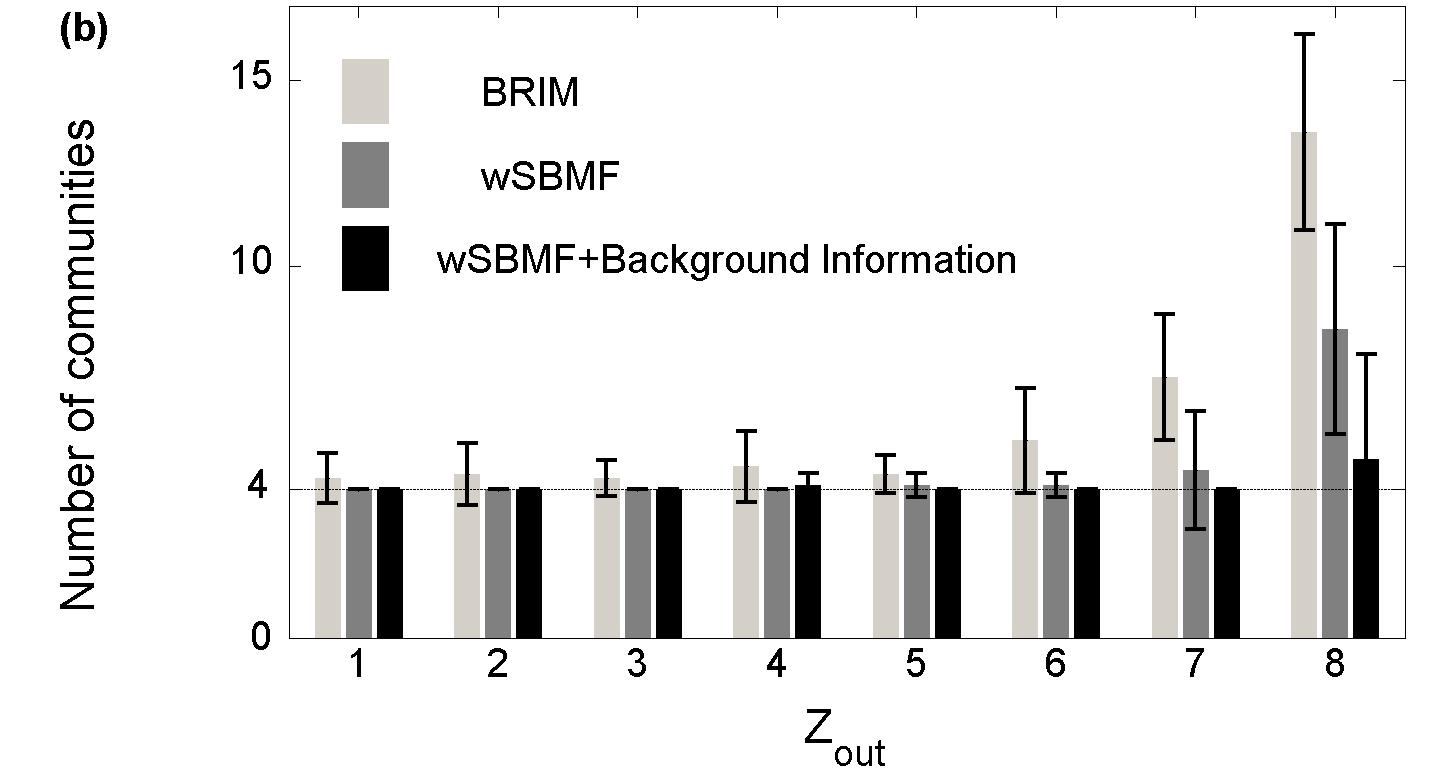}
\end{center}
\caption{{\bf Performance of BRIM and wSBMF on the bipartite networks, SBMF on the monopartite networks, and the number of communities estimated
by BRIM and wSBMF on non-overlapping networks.} We randomly select five percent of pairs in $\Delta$ for background information.}
\label{Fig:02}
\end{figure}

\begin{figure}[!ht]
\begin{center}
\includegraphics[height=50mm,width=70mm]{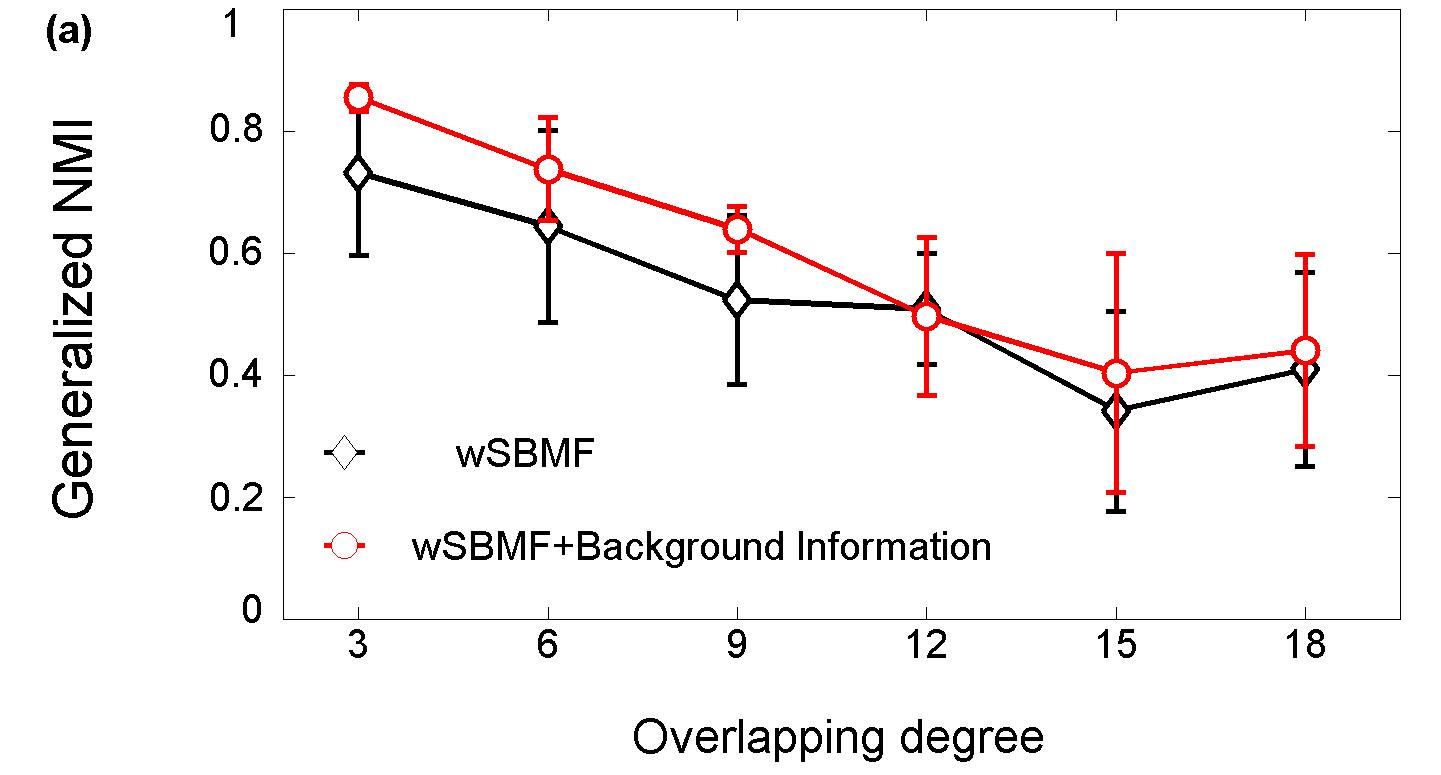}\vspace{5mm}
\includegraphics[height=50mm,width=70mm]{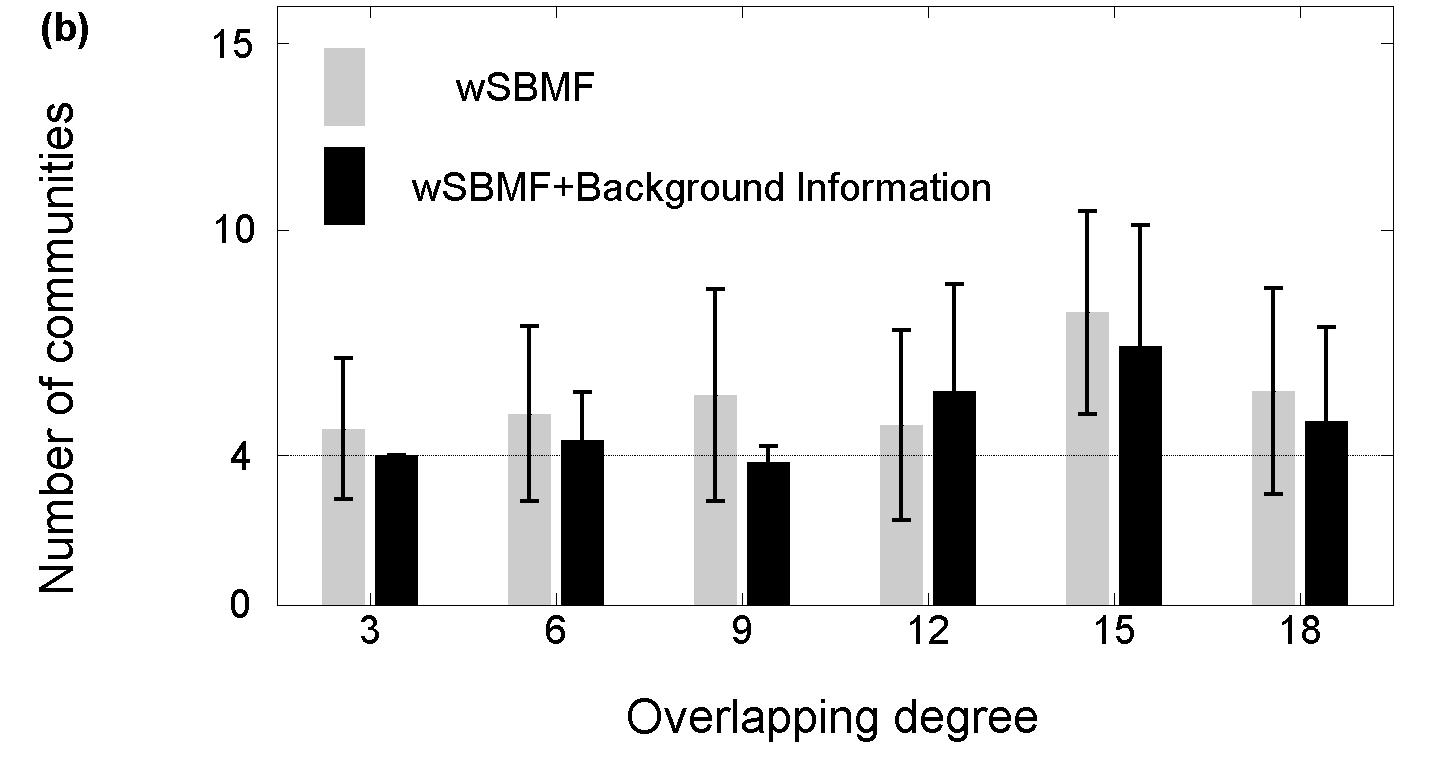}
\end{center}
 \caption{{\bf Performance of wSBMF and the number of communities estimated
by SBMF on overlapping networks.} We randomly select five percent of pairs in $\Delta$ for background information.}\label{Fig:03}
\end{figure}

\begin{figure}[!ht]
\begin{center}
\includegraphics[height=37mm,width=80mm]{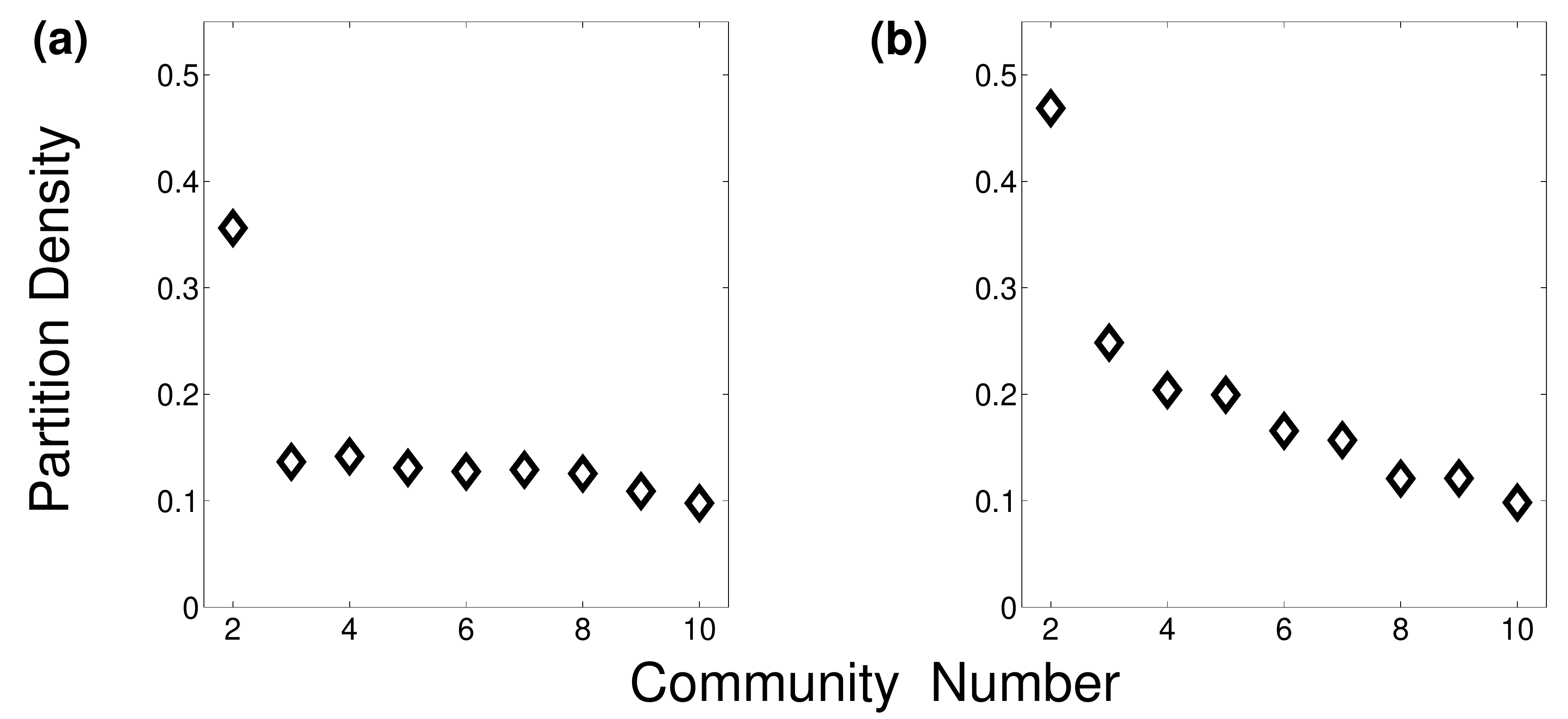}
\end{center}
 \caption{{\bf Averaged partition density of wSBMF versus community number on (a) women network and (b) senator network.}}\label{Fig:04}
\end{figure}

\begin{figure}[!ht]
\begin{center}
\includegraphics[height=40mm,width=70mm]{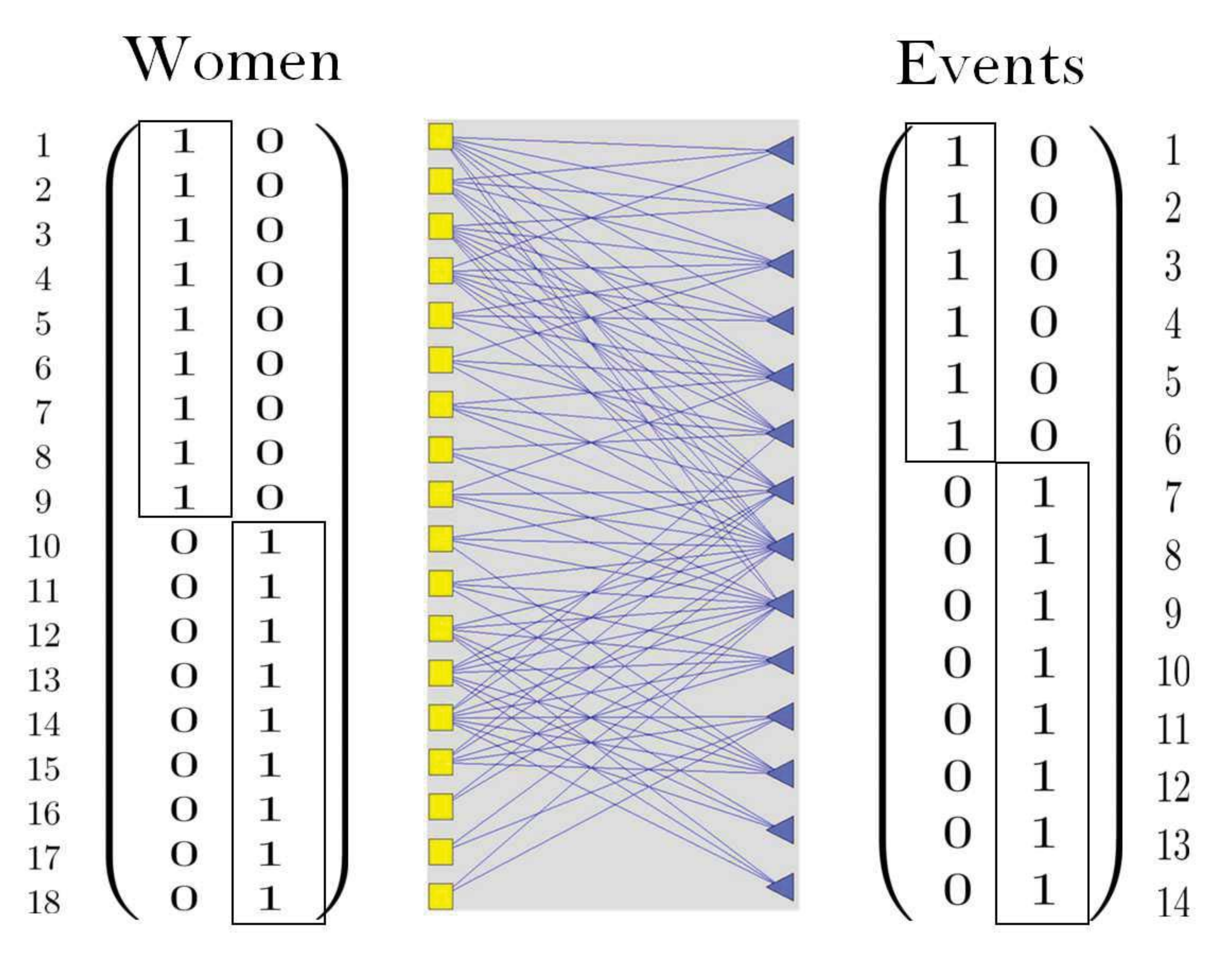}
\end{center}
 \caption{{\bf Communities detected by wSBMF model in the women network.} There are no
outliers and overlapping nodes.}\label{Fig:05}
\end{figure}

\begin{figure}[!ht]
\begin{center}
\includegraphics[height=45mm,width=55mm]{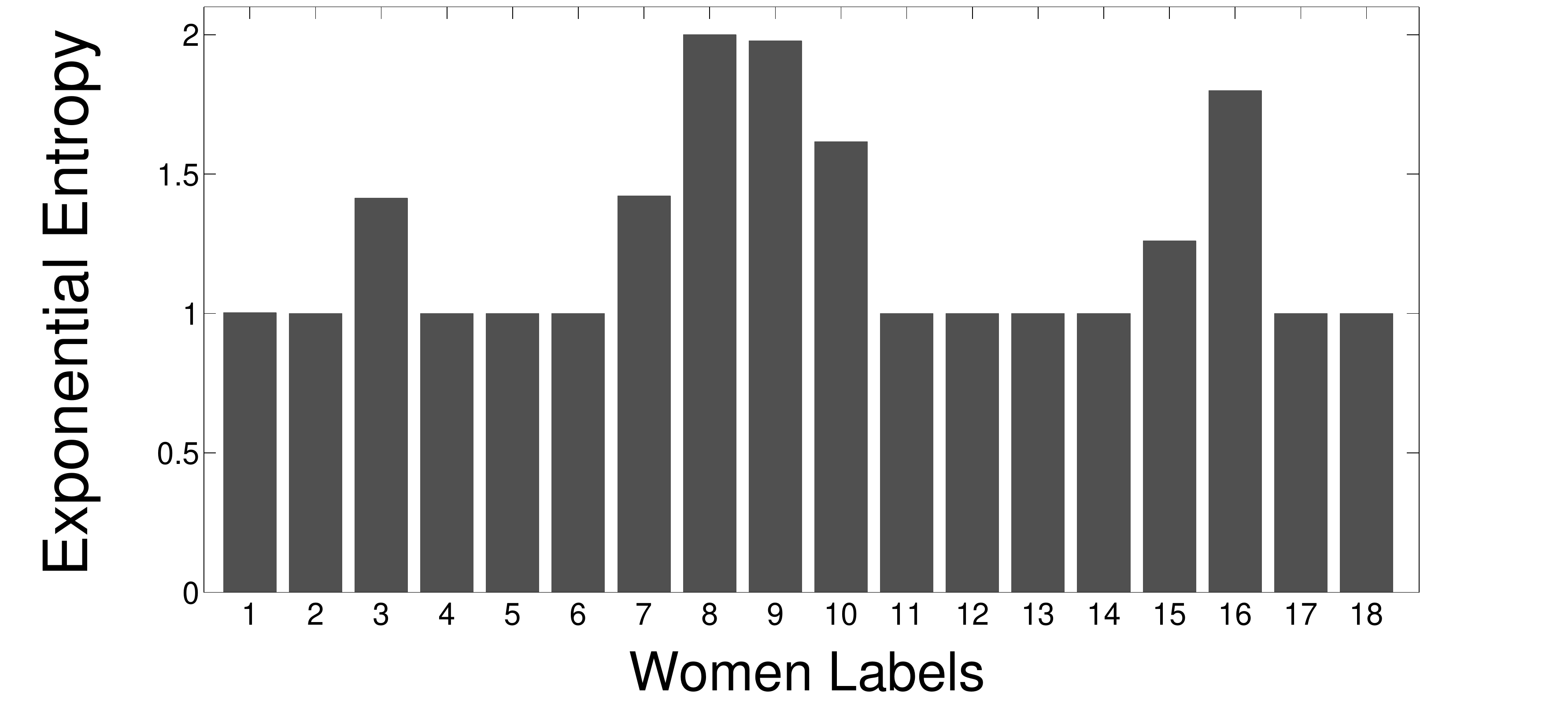}
\end{center}
 \caption{{\bf Exponential entropy of women.} Higher value means fuzzier membership degree.}\label{Fig:06}
\end{figure}

The results are shown in
Figs. \ref{Fig:02} and \ref{Fig:03}. They show that the wSBMF method is much better than SBMF on unipartite networks, indicating the nonreducible property of community detection problem in bipartite networks, and it also performs
better than BRIM in non-overlapping community benchmark graphs. Our method can
identify reasonable number of communities, and the background information can
significantly improve the results.
We also evaluate the method on the southern women network and the senator
network. Fig. \ref{Fig:04} shows the results of partition density under
different community numbers on the two networks, and the most appropriate
number is 2 for both of them. For the southern women network, the result is very similar
to that in~\cite{davis}, where there are two groups in women,
women $1-9$ and $9-18$. For the senator network, the result is consistent
with  American two-party politics. Fig. \ref{Fig:05} shows the result of
community structure on the women network detected by wSBMF. We also use
exponential entropy $e^{H_i}, i=1,2,\ldots,n_\Delta$ \cite{entropy}, to analyze
the strength of women's community memberships, where $$ H_i =
-\sum_{j=1}^2U_{ij}\log U_{ij}, i=1,2,\ldots,n_\Delta.  $$ The result is given
in Fig. \ref{Fig:06}.

\section{Discussion}
In this paper we have shown how to apply symmetric binary matrix factorization and partition density to find communities in bipartite networks. The model is parameter free, easy to implement, and flexible enough to incorporate background information. Experimental results on both the synthetic and  real-world networks demonstrate the effectiveness of the proposed method.

There are two interesting problems for future work: (i) extension of the method to weighted bipartite networks and directed bipartite networks; and (ii) theoretical investigation on partition density and algorithm design for its direct optimization.

\section *{Appendix}
Summarization of Algorithm \ref{Al:01} and \ref{Al:02}. We set the iteration number $\mbox{C}_1$ equal to 10 and the iteration number $\mbox{C}_2$ equal to 100.
 \begin{algorithm}
 \caption{Nonnegative Matrix Factorization
  (Alternative Least Squares Error)}
 \label{Al:01}
 \begin{algorithmic}[1]
 \REQUIRE $B, \mbox{C}_1$
 \ENSURE $U_1, U_2$
 \STATE Initialize elements of $U_1$ with nonnegative random  numbers drawn from $[0,1]$.
 \FOR{$t=1:\mbox{C}_1$}\vspace{2mm}
 \STATE Solve for $U_2$ in equation
 $
  U_1^TU_1U_2=U_1^TA
 $\vspace{2mm}
 \STATE $U_2=\max(U_2,0)$
 \STATE
 Solve for $U_2$ in equation
 $
  U_2U_2^TU_1^T=U_2A^T
 $\vspace{2mm}
 \STATE $U_1=\max(U_1,0)$
 \ENDFOR
 \end{algorithmic}
 \end{algorithm}
 \begin{algorithm}[H]
 \caption{Weighted Symmetric Nonnegative Matrix Factorization
  (Multiplicative Updates)}
 \label{Al:02}
 \begin{algorithmic}[1]
 \REQUIRE $A,U, \mbox{C}_2$
 \ENSURE $U$
 \FOR{$t=1:\mbox{C}_2$}\vspace{2mm}
 \STATE
 $
 \displaystyle U:=U\circ\frac{\left[(L\circ A)U\right]}{\left[L\circ (UU^{T})U\right]}
 $\vspace{2mm}
 \ENDFOR\vspace{2mm}
 \STATE
 $
 \displaystyle{U_{ij}:=\frac{U_{ij}}{\sum_jU_{ij}}}, i\, = 1,2,\cdots,n
 $
 \end{algorithmic}
 \end{algorithm}

\bibliographystyle{cpb}
\bibliography{bipartite}

\end{document}